\documentclass[aps,twocolumn,showpacs,preprintnumbers,amsmath,amsfonts,mathtools,array,amssymb,amsthm,revsymb,prb,superscriptaddress,
longbibliography]{revtex4-1}
\usepackage{graphicx}
\usepackage{txfonts}
\usepackage{mathptmx}
\usepackage{braket}
\usepackage{color}
\pdfmapfile{+txfonts.map}

\begin{document}

\title{Modelization of charge carriers mobilities in halide perovskites : \\  Fr\"ohlich scattering and quantum  localization effects in a dynamic disorder regime}

\date{\today}

\author{Antoine Lacroix} \email{antoine.lacroix@neel.cnrs.fr}
\affiliation{Universit\'e Grenoble Alpes, CNRS, Institut NEEL, F-38042 Grenoble, France}
\author{Guy Trambly de Laissardi\`ere}  \email{guy.trambly@u-cergy.fr}
\affiliation{Laboratoire de Physique Th\'eorique et Mod\'elisation, CNRS and
Universit\'e de Cergy-Pontoise,  95302 Cergy-Pontoise, France}
\author{Pascal Qu\'emerais}  \email{pascal.quemerais@neel.cnrs.fr}
\affiliation{Universit\'e Grenoble Alpes, CNRS, Institut NEEL, F-38042 Grenoble, France}
\author{Jean-Pierre Julien}  \email{jean-pierre.julien@neel.cnrs.fr}
\author{Didier Mayou }  \email{didier.mayou@neel.cnrs.fr}
\affiliation{Universit\'e Grenoble Alpes, CNRS, Institut NEEL, F-38042 Grenoble, France}

\date{\today{}}

\begin{abstract}
{We analyze the quantum transport properties of MAPbI3 within a tight-binding model. Charge carriers are strongly scattered by the Fr\"ohlich interaction with longitudinal optical phonon modes. This limits their mobilities at room temperature to the order of 200 cm$^2$/Vs. In the presence of additional extrinsic disorder the mobility decreases and a large fraction of the electronic states at band edges  can be localized. These states would be insulating if the lattice were static, but their localization is broken by the dynamic disorder induced by the vibrations of the longitudinal optical modes. This process of electrons and holes diffusion, driven by the lattice dynamics, contributes  to the unique electronic properties of this material.}

\end{abstract}

        \maketitle
	\pagenumbering{arabic}

	Metal halide perovskite have recently emerged as great materials for photovoltaic and optoelectronic devices\cite{Green2014,Intriguing_Properties}. The rapid progresses obtained in these applications require also to gain a better fundamental understanding of their electronic properties. Therefore there is a need for improved theoretical models in particular for the electronic transport , and so far there is no consensus on a detailed theory of charge transport in these materials \cite{theory, advances}. Owing to their softness, the very low phonons frequency and the evidence for anharmonicity a strong dynamic thermal disorder  develops at room temperature\cite{Anh, born, dynamic_disorder_perovskite}. For a perfect crystal at room temperature  theoretical investigations point toward the importance of the  scattering of charge carriers by longitudinal optical (LO) phonon modes \cite{Herz}. This coupling to LO modes could also lead to the formation of large polarons. Although the mass renormalization estimated to about 40 percent is moderate this polaronic effect is often considered in the literature \cite{Polaron_stabilization,Polaron_radius, Polaronic_energies}. The dipolar moment of Methylammonium (MA) has also been considered as a possible source of scattering even though recent calculations suggest that the scattering by the associated dipolar field has a limited effect on transport \cite{mu_MA}. Other source of scattering have also been considered related to an anharmonic behavior of these crystals at room temperature \cite{anharmonicity_lifetimes}. Extrinsic disorder is of course present in real systems but its effect is also difficult to model.

In this letter we investigate the transport properties of electrons and holes in MAPbI$_{3}$ (MAPI) within a tight binding model which allows to perform all the calculations of electronic transport in real space \cite{1, 2,3,4}. We take into account the effect of intrinsic thermal disorder of the PbI$_3$ matrix as well as the  dipolar field created by the MA cation. We use the recently introduced analytical Drude-Anderson model \cite{Drude_Anderson}  which is useful to interpret numerical or experimental results. From this model we extract the basic characteristics of the quantum transport such as scattering time, velocity correlation and backscattering effect which are at the heart of quantum localization. These results show that the polaronic state is not stable at room temperature and that the mobilities are determined mainly by the Fr{\"o}hlich  scattering and by quantum localization effects in a dynamic disorder regime. This dynamic disorder tends to break the quantum localization by dephasing which allows charge carriers diffusion. This process of diffusion is neither a band like process nor a thermally activated hoping and contributes to the unique properties of this material.

	We use a tight binding model approximation to describe our material. The Hamiltonian is written as:
\begin{equation}
\begin{gathered}
   \hat{H} = \sum_{i \mu \sigma}  \ket{i \mu \sigma} \epsilon_{i \mu} \bra{i \mu \sigma} 
   + \sum_{i \mu,j \mu', \sigma} \ket{i \mu \sigma} t_{i \mu ,j \mu' } \bra{j \mu' \sigma} \\
    + \sum_{i \mu \sigma, \mu' \sigma'} \ket{i \mu \sigma}  \lambda_{i \mu \sigma,\mu' \sigma' }  \bra{i \mu' \sigma'},
\end{gathered}    
\end{equation}
where $\epsilon_{i \mu}$ is the onsite energy, of orbital $\mu$ on atom $i$, $t_{i \mu, ,j \mu' }$ is the hopping integral between orbitals $\mu$ on atom $i$ and orbital $\mu'$ on a neighboring atom $j$, and $\lambda_{i \mu \sigma,\mu' \sigma' }$ is the spin orbit coupling between orbitals of the same atom, but of different spin $\sigma$ and $\sigma'$. The orbitals participating in transport are the valence orbitals of each elements which are of s and p type. The MA molecule energy levels are too high compared to Lead and Iodine to participate in charge transport, and do not appear in the tight binding model, but they provide one electron per unit cell to the system. This type of Hamiltonian is very well suited for numerical studies. It can be separated in three parts, the diagonal elements, the off-diagonal elements, and the spin orbit elements. The tight binding parameters of the system without disorder have been fitted \cite{even} to be consistent with MAPI's $\it{ab-initio}$ band structure calculations \cite{BrivioSchil-fgaardPRB2014,TowfiqJXZEPL2014} , and are used as a basis for this work.

We consider now the description of the intrinsic thermal disorder. The spin orbit coupling has a strong impact on MAPI's band structure, due to the splitting of the Lead p orbitals, but it is a local phenomenon and we assume that it is not affected by thermal disorder. So we only have to account for off-diagonal and diagonal disorder. MAPI's Debye temperature is around 175K \cite{Debye}, much lower than room temperature, and we adopt a classical description for phonons since we focus mainly on room temperature properties.

The off-diagonal disorder is related to the distance and overlapping between orbitals. We describe the displacement of atoms using independent Einstein's oscillators. In this model, all atomic displacements are statistically independent, and the  potential is treated as harmonic and isotropic, even though the environment of $I$ is not of cubic symmetry. Such single particle potentials have been computed from experimental data by Tyson et al. \cite{tyson}. For such a model, the distribution of atomic displacements is a gaussian centered on zero and of standard deviation $\sqrt{k_bT}/(\omega_0\sqrt{m})$ where  $\omega_0$ is  the pulsation, $m$ is the atom's mass and $k_bT$ is the thermal energy. For Lead ($Pb$)  $\hbar \omega_0 = 6.4$ meV and for Iodine ($I$) $\hbar \omega_0 = 5.4 $meV. Once an atomic configuration is set the hopping parameters are recomputed from the Slater-Koster relations \cite{Slater} for changes in direction, and via a power law in $(d_0/d)^2$ for changes in distance from $d_0$ to the actual distance $d$. 

The disorder of the onsite energies is determined by the electrical potential inside the material. The variation of this electrical potential with respect to the periodic structure is caused by the displacement of the charged ions and the orientation of the MA molecules. The contribution of the MA molecules to the diagonal disorder is found small \cite{mu_MA} (see supplementary material), and has a negligible effect on mobilities. Therefore we present results only with diagonal disorder due to the displacement of Lead and Iodine ions. The displacement of an atom creates a dipole moment which is the product of the displacement by the Born charge in the considered direction. We use the Born charges computed by Pérez-Osorio et al.\cite{born} for the Lead and Iodine atoms. The main contribution to the electrostatic potential is expected to come  from  the LO phonon modes, which energies are in the 10-13 meV range\cite{Herz}. Here we also represent the statistics of the displacements by using an Einstein model of independent isolated atomic oscillators with an effective phonon pulsation $\omega_E$ close to $10$ meV.  We note that the phonon modes that contribute most to the diagonal disorder are the LO modes, which represent only a small portion of the complete set of modes\cite{Phonon_spectra,Herz}, while the off diagonal disorder gets contribution from all modes. We thus assume that both disorders are statistically independent. 

From  the relation between the dipolar moments $\vec{d_{i}}$, $k_bT$ and $\omega_E$, we find (see supplementary material) that the probability distribution of the dipoles is a Gaussian and that the variable $W= k_bT/(\epsilon_r \omega_E)^2$ is the parameter that determines the electrostatic potential statistics due to thermal disorder.  In the rest of the paper we fix the temperature to $T= 300K$, the relative dielectric constant to $\epsilon_r=5$\cite{dielectric}. We find that the value of $\omega_E$ that reproduces best the gap at room temperature  is close to $10$ meV (see supplementary material) and discuss only the effect of varying  $\omega_E$ around $10$ meV. $W$ increases when disorder increases, meaning that a decrease of $\omega_E$  corresponds to an increase of disorder. Finally we note that the exact structure and type of defects in MAPbI3 are not fully understood and some types of disorder that are not described in the present harmonic model could play a role. Yet the central idea of this work is that the transport properties are mainly determined by a single parameter,  that describes the strength of disorder and the scattering rate. Here this parameter is $\omega_E$ (or $W$).

\begin{figure}
\includegraphics[width=0.99\linewidth]{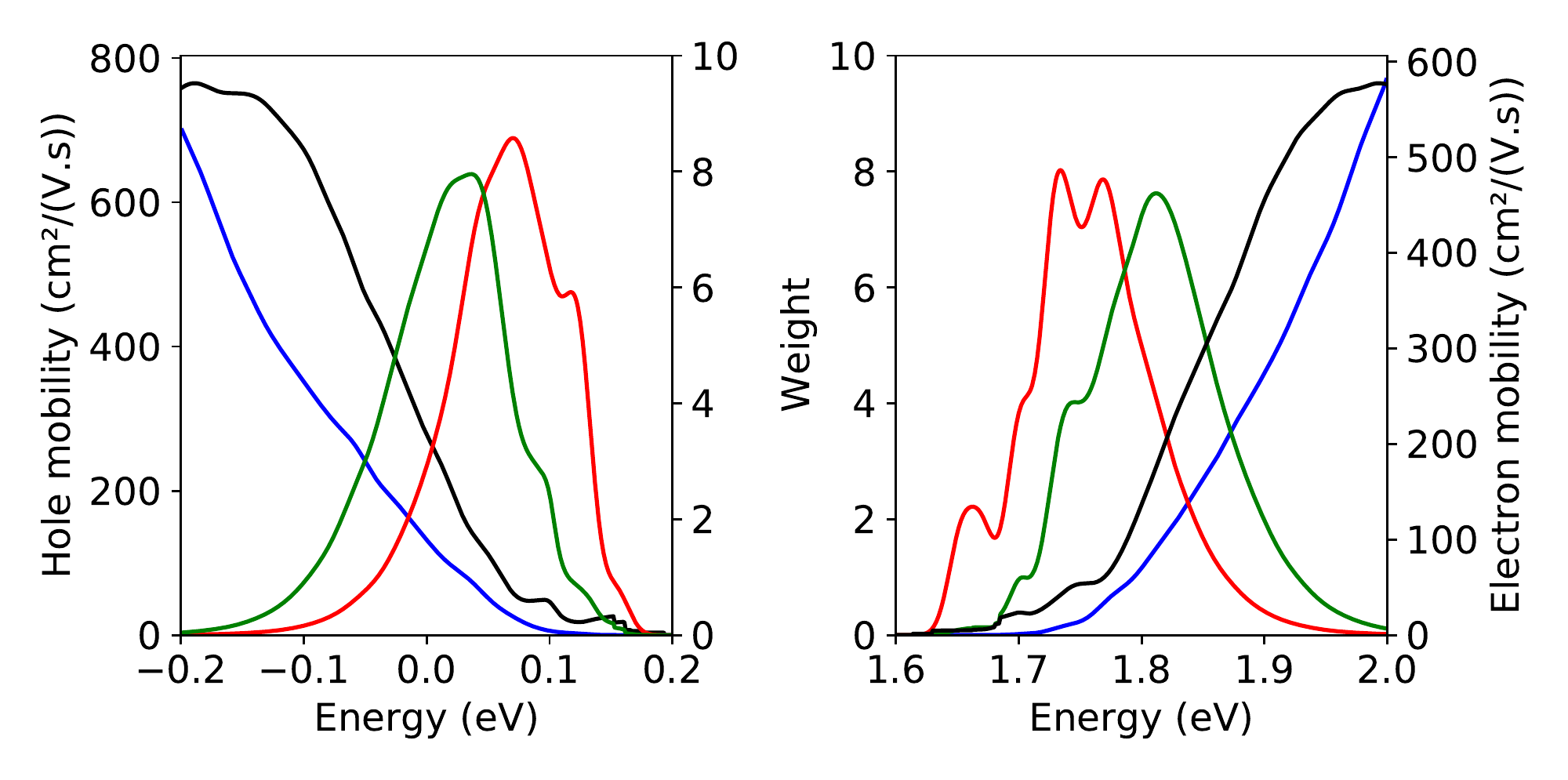}
\caption{(Color online) Variations of the  carrier mobilities (black), the density of states (blue), the occupied DOS (red) and the differential conductivity (green) as a function of energy, near the band edges, for holes \textbf{(left)} and electrons \textbf{(right)}. Except for the mobility  the quantities are normalized (for each type of carrier) such that their integral is one on the represented energy interval  (axis labeled weight) }
\label{fig:DOS}
\label{DOS}
\end{figure}

We first present results for $\omega_E =10$ meV showing  energy resolved analysis for the properties of electrons and holes. Figure \ref{fig:DOS} shows the  density of states (DOS) $n(E)$, the occupied density of states $n(E) exp(-\beta E)$, the mobility $\mu(E)$ at energy $E$   and the differential conductivity  $d\sigma/dE = \mu(E) n(E) exp(-\beta E)$. We find that the occupied DOS is important within a range of about $0.2 eV$ and that states contributing the most to the transport of current are also within a range of about $0.2 eV$ but slightly shifted away from the band edges. The very small mobility close to the band edge  is related to the  Urbach's tail, but it cannot be analyzed with the present energy resolution that is limited to $10-20$ meV. The typical value of the mobility for states that contribute most is in the range 100-200 cm$^2$/Vs which corresponds indeed to the best experimental mobilities. We note that results for electrons and holes are similar and in the rest of the paper we discuss only averaged properties of both charge carriers. Finally we note that the densities of states of both electrons and holes are very different from the parabolic shape typical of free particule with an effective mass. This indicates that the disorder in this model is strong and that the energy broadening of a state due to the disorder is comparable to its energy, counted from the band edge. According to the Ioffe-Regel criterion\cite{Ioffe-Regel} one expects that quantum localization effects are important and we analyze now more precisely these quantum effects.

We start our discussion of the quantum localization effects  by considering the  Drude-Anderson model \cite{Drude_Anderson} that was introduced recently and that fits very well the results of the present numerical study (see supplementary material for the fitted parameters). Two  fundamental quantities for the quantum transport are the mean squared displacement $X^2(t)$ (that is computed through the RSKG method)  and the velocity correlation function $C(t)$. They are defined for each type of carrier and are related by :
	\begin{equation}
\frac{1}{2} \frac{d X^2(t)}{dt}=  \int_0^{t} C(t')dt',
\end{equation}
and the optical terahertz conductivity, $\sigma(\omega)$, (i.e. the real part of the full complex conductivity) can also be computed from $C(t)$ or $X^2(t)$ \cite{Drude_Anderson, Map_mobility}. Using the Kubo formalism one can derive : 
\begin{equation}
\sigma(\omega)=  e^2 n \frac{tanh(\beta\hbar\omega/2)}{\hbar\omega/2} Re\int_0^{\infty} e^{i\omega t} C(t) dt ,
\end{equation}
where $e$ and $n$ are the charge and concentration for each carrier type. Thus the optical terahertz conductivity, contains informations about the charge carriers dynamics and the temporal behavior of the spreading of electronic states $X^2(t)$. 

The Drude-Anderson model consists in assuming  a phenomenological expression for the velocity correlation $C(t)$ :
\begin{equation}
C(t)  \simeq   C_C e^{-t/\tau_C} - C_B e^{-t/\tau_{B}}e^{-t/\tau_{\Phi}}.
\label{velocity correlation}
\end{equation}
The first term on the right corresponds to the standard classical picture of electronic transport where scattering events lead to a loss of the memory of the initial velocity on a characteristic time $\tau_C$. The second term on the right represents a negative contribution to the velocity correlation function. This corresponds to the backscattering phenomena which is at the heart of the Anderson localization phenomenon. Physically one must have $\tau_{B} > \tau_C$ because the backscattering phenomena occurs after several scattering events. The term $e^{-t/\tau_{\Phi}}$ describes an exponential damping of the backscattering terms due to dephasing processes that do not conserve the charge carrier energy. In the present case the dephasing of the backscattering terms is due to the dynamics of the lattice and $\tau_{\Phi}$ is of the order of the period of the LO modes, which are the main source of scattering, i.e. $\tau_{\Phi} \simeq 1/\omega_{E}$. In the following we shall discuss the results by considering mainly the time dependent and frequency dependent  mobilities defined by : 
\begin{equation}
\tilde {\mu}(t) = \frac{e}{k_b T} \frac{X^2(t)}{2t}, \quad  \tilde{\mu}(\omega) = \frac{\sigma(\omega)}{n e },
\end{equation}
where $e>0$ is the electron charge and the mobility $\mu$ is given by  $\mu = \tilde {\mu}(t \rightarrow \infty) = \tilde{\mu}(\omega \rightarrow 0)$. We define also the static mobility $\mu_{S}$ and the classical mobility $\mu_{C}$ by :
\begin{equation}
\mu_{S}= \frac{e}{k_b T} (C_C \tau_C - C_B \tau_{B}) ,\quad \mu_{C}= \frac{e}{k_b T} (C_C \tau_C).
\label{SC}
\end{equation}
In the absence of dephasing i.e. by considering  the  lattice as static  ($\tau_{\Phi} \rightarrow \infty$) one has $\mu=\mu_{S}$, and the ratio $R= \mu_{S}/ \mu_{C}$ is an indicator of the importance of localization effects.
\begin{figure}
\includegraphics[width=\linewidth]{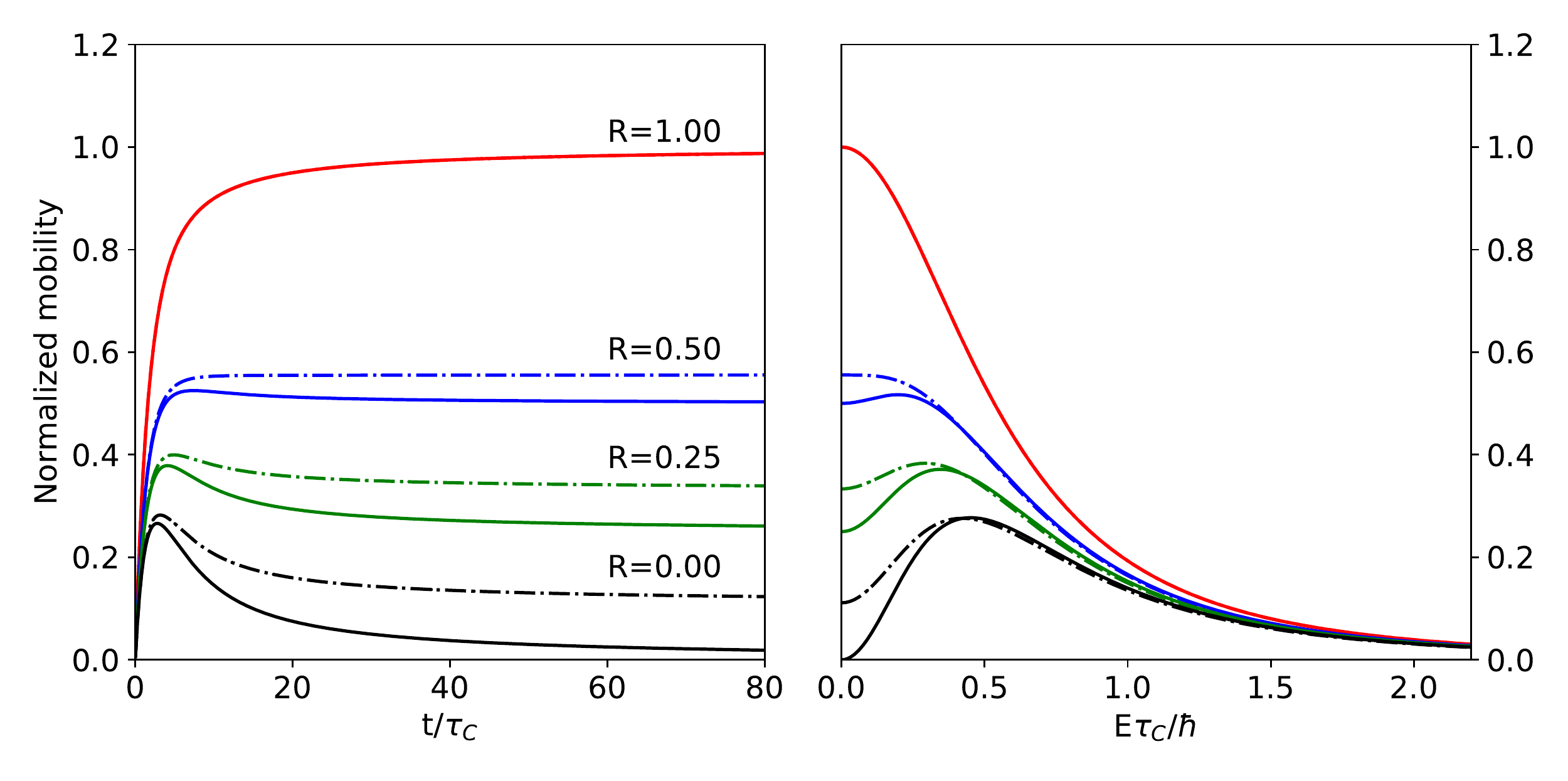}
\caption{(Color online) Normalized time dependent mobility $\tilde {\mu}(t)/\mu_{C}$  \textbf{(left)} and normalized frequency dependent mobility   $\tilde{\mu}(\omega)/\mu_{C}$ \textbf{(right)} for different localization ratio $R= \mu_{S}/ \mu_{C}$. The full line corresponds to results without dephasing processes and the dashed line corresponds to results with a dephasing process. Here $\tau_{B} = 2.5 \tau_{C}$ and  $\tau_{\phi} = 8\tau_{B}$ .}
\label{regimes}
\end{figure}
Figure ($\ref{regimes}$) shows the different regimes of diffusion and the corresponding optical conductivities. We consider typical  regimes of diffusion : one which is classical and presents no localization effect ($R \simeq 1$), one which is strongly localized with ($R \simeq 0$), and  two intermediate cases. At short time there is always a ballistic regime during which the charge carrier moves through the material without interacting with disorder i.e. $ \tilde {\mu}(t) \propto t $. The end of this regime marks the elastic diffusion mean free time. This ballistic regime can be followed by a quantum diffusion regime which happens when disorder becomes strong enough so that charge carriers become localized, resulting in a drop in diffusivity and in mobility. Finally at sufficiently large time the diffusive regime is reached for which $ X^2(t) \propto t $ and the diffusivity and the time dependent mobility $\tilde {\mu}(t) $ are constant. For a completely localized system $\tilde {\mu}(t) $ tends exactly to zero at large times. Finally we emphasize that when the quantum localization increases the Drude peak is progressively replaced by a dip in the frequency dependent conductivity (figure ($\ref{regimes}$) right panel).

In the presence  of dephasing processes (i.e. for a finite $\tau_{\Phi}$), the mobility $\mu$ is the sum of a static contribution $\mu_{S}$ and of a contribution due to the lattice dynamics $\mu_{LD}$ with : 
\begin{equation}
\mu=\mu_{S}+\mu_{LD} ,  \quad  \mu_{LD}= \frac{e}{k_b T}  \frac{L^2(\tau_{\Phi})}{2 \tau_{\Phi}}.
\label{LD}
\end{equation}
Since the dynamics of the lattice tends to break the effect of quantum localization $\mu_{LD}$ is positive with $L^{2}(\tau_{\Phi})=2C_B \tau_{B}^2/(1+\tau_{B}/\tau_{\Phi})$. In the classical  limit there is no backscattering $C_B=0$ and $\mu_{LD}=0$. In the weak-localization regime there is some backscattering, and $\mu_{LD}$  gives  a correction to the static  mobility $\mu_{S}$. In the strongly localized limit $\mu_{S} \simeq 0$ but transport is still possible thanks to the $\mu_{LD}>0$ term, and in this limit $L^2(\tau_{\Phi})$ is essentially equal to the localization length (see supplementary material). The physical picture is that the charge can diffuse up to a maximum extent which is the localization length. Then the diffusion stops and can start again only  if a dephasing process breaks the electronic coherence. In this regime the diffusion coefficient is inversely proportional to the dephasing time. Thus the diffusion  is driven by the lattice dynamics and indeed  the diffusion coefficient increases linearly with the vibration frequency $\omega_{E} \propto 1/\tau_{\phi}$. Let us emphasize that this regime where diffusion results from a dephasing process differs from the hopping regime at low temperature that is thermally activated. We note also that in this regime of lattice driven diffusion the mobility decreases when the temperature increases, just as in a band like conduction regime. Indeed increasing temperature and therefore static disorder decreases the localization length and  $L^2(\tau_{\Phi})$ without changing $\tau_{\Phi}$. From equation \ref{LD} this decreases $\mu_{LD}$. The physical picture of the dephasing process corresponds to the so-called Thouless regime \cite{Thouless1,Thouless2,Thouless3,Thouless4} that is expected to occur near a metal-insulator transition. It  is  also equivalent to the concept of transient localization that has been proposed for crystalline organic semi-conductors like rubrene \cite{Transient_Loc, Fratini_2016, rubrene}.

We come now to the study of charge carriers mobilities. By applying  the RSKG (Real Space Kubo-Greenwood)\cite{1, 2,3,4} method we perform a study  of quantum transport without resorting to perturbative treatment of disorder. This method has been applied so far only to systems of one and two dimensions. The present calculation is the first application to a three dimensional system where we use systems containing several millions of unit cells.  This opens new perspectives for the use of the RSKG method. Since holes and electrons have similar mobilities (see figure \ref{fig:DOS}) we present values of the time dependent mobilities $\tilde {\mu}(t)$ and optical terahertz conductivities averaged over electrons and holes. The optical conductivity reflects the dynamics of the charge diffusion and could bring much information, but experimental results obtained so far appear somewhat contradictory\cite{Terahertz,THZ_high}. Further experimental studies are needed and could be compared to the present results.

The model with $\omega_E=10$ meV is the one expected to be the closest to perfect bulk MAPI, $\omega_E=12.5$meV serving as a high expectation value, and $\omega_E=7.5$meV being used to describe a more imperfect MAPI sample with extrinsic disorder. The Drude-Anderson model fits the data of $\tilde {\mu}(t)$ very accurately which allows to derive its parameters from the numerical calculation of $X^2(t)$. We find that the LO phonon modes of MAPI limit the mobilities to maximum values of about 200 $cm^2/(V.s)$ with  $\mu_{S}$ and $\mu_{LD}$ contributions that are nearly equal. Even in this case of relatively high mobility  quantum localization effects are strong with $R \simeq 0.15$ . The elastic mean free path $l= \sqrt{\Delta X^2(\tau_C)}$ is about  20 $ \mathring{A}$ (resp. 30 $ \mathring{A}$) for $\omega_E=10$ meV (resp. $\omega_E=12.5$ meV). The scattering times $\tau_C$ are about $8.3$  (resp $10.3$) Femtoseconds  for $\omega_E=10$ meV (resp. $\omega_E=12.5$ meV). As expected the backscattering times $\tau_B$ are 2-3 times larger than $\tau_C$ (see supplementary material).

\begin{figure}
\includegraphics[width=\linewidth, clip]{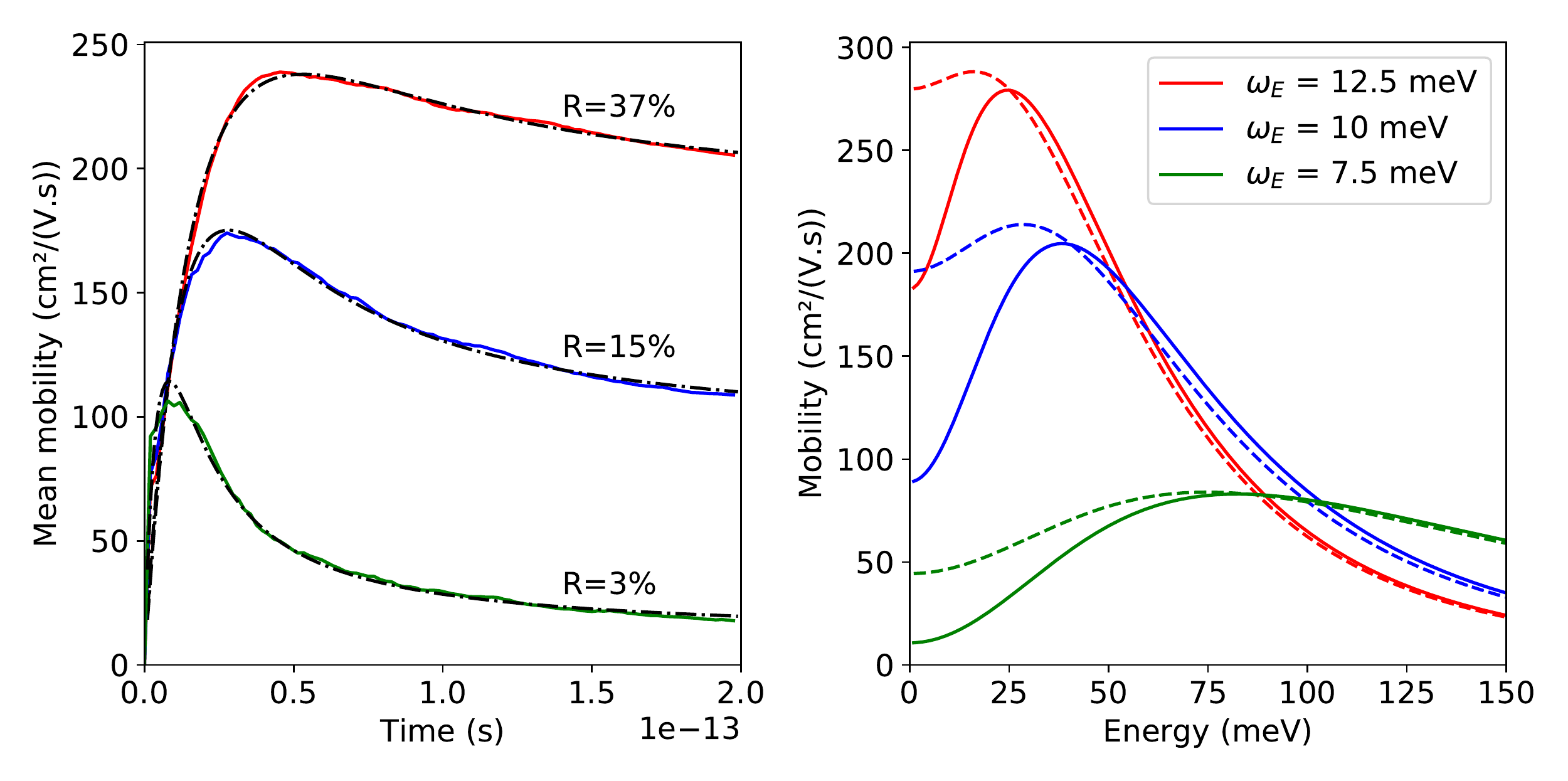}
\caption{(Color online) Mobilities averaged over hole and electron states. \textbf{Left}  Mobility as a function of time for different values of $\omega_{E}$ \textit{(full line)}, and the corresponding fits using the Drude-Anderson model \textit{(black dashed line)}. The ratio $R= \mu_{S}/ \mu_{C}$ is obtained from the fitted parameters. \textbf{Right} Mobilities as a function of frequency calculated from the Drude-Anderson model. Without dephasing processes \textit{(full line)}  and with dephasing \textit{(dotted line)}.The zero energy mobility is equal to $\mu_S$ (without dephasing) or $\mu_S + \mu_{LD}$ (with dephasing).The dephasing time is $\tau_{\phi}=7.10^{-14}s$.}
\label{liste_mob}
\end{figure}

The results for $\omega_E=7.5$ meV show the quick increase of the quantum localization with additional disorder ($R\simeq 0.03$). The scattering time is $\tau_C\simeq 2.6$ Femtoseconds and a  large portion of the charge carriers states close to the band edges are strongly localized on a  time scale $\tau_B \simeq 8$ Femtoseconds, which is short compared to the phonon period (of about $400$ Femtoseconds). Figure (\ref{liste_mob}) shows that for this disorder the mobility is of the order of $\mu \simeq 50 cm^2/V.s$ with $\mu_{S}\simeq 10 cm^2/V.s$ and $\mu_{LD}\simeq 40 cm^2/V.s$.Therefore in this regime the mobility is dominated by the contribution due to the lattice dynamics. We note that the RSKG method tends to underestimate the localization so that for these parameters the total mobility should have an even lower relative contribution from the static mobility. 

These results allow to discuss the formation of large polarons at room temperature. The coupling constant $\alpha \simeq 2-3$ implies the formation of large polarons at low temperatures.  The formation energy of the  polaron is given by  $E_{Pol} \simeq \alpha \hbar \omega_E \simeq 30$ meV \cite{Polaron_radius, Polaronic_energies}. Yet in the high temperature limit, i.e. above the Debye temperature (175 K in MAPI \cite{Debye}), the phonon modes are thermally excited and this induces a scattering of electrons and a tendency to erase the polaronic state. This is shown for example by recent theoretical calculations in the case of CsPbBr3\cite{Bernardi}. This scattering remains delicate to compute with standard polaron theories\cite{PEETERS198481, Calculating_polaron_mobility}.The scattering by thermal vibrations induces an energy broadening  $\Delta E \simeq \hbar /\tau$ where $\tau$ is the electron lifetime due to scattering. In our case due to the long range nature of the scattering by longitudinal modes there is forward scattering and  $\tau \leq \tau_C$. Then  for the systems with highest mobility $\mu \simeq 200 cm^2/Vs$ the energy broadening is $\Delta E \geq \hbar /\tau_C\geq 90$ meV. Therefore even for this less disordered cases the energy broadening due to the {Fr\"ohlich} scattering  is larger than the polaron formation energy  $\Delta E> E_{Pol} $. Correlatively the estimated polaron radius of about  40-50 $ \mathring{A}$ is larger than the elastic mean free-path which is less than 20-30 $ \mathring{A}$. This indicates that because of the moderate coupling constant $\alpha \simeq 2-3$  the polaron state is erased by the strong disorder  that exists well above the Debye temperature, at room temperature. This justifies the present model that consist, in a first step, to neglect at high  temperature the action of the charge carrier on the lattice and just retain the action of the lattice on the charge carrier.

\begin{figure}
\includegraphics[width=0.99\linewidth]{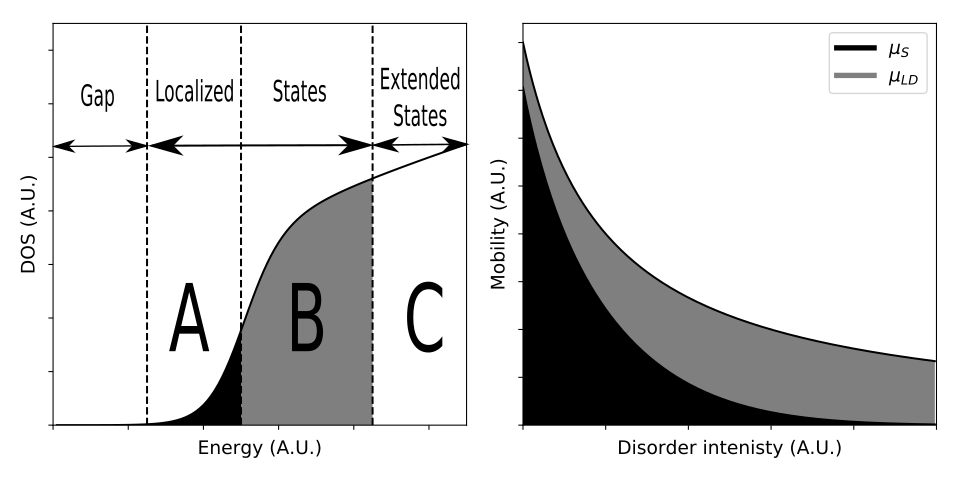}
\caption{\textbf{Left:} Schematic representation of the different type of states as a function of energy. Region A : diffusion dominated by thermally activated hopping. Region B : diffusion is through  the dephasing  processes induced by the lattice dynamics. Region C : band-like  diffusion. \textbf{Right:} Schematic representation of the evolution of the mobility $\mu=\mu_{S}+\mu_{LD}$  as a function of disorder. Static contribution  $\mu_S$ \textit{(black)} and  lattice driven mobility $\mu_{LD}$ \textit{(grey)} are shown. }
\label{Scenario}
\end{figure}

Figure (\ref{Scenario}) summarizes the scenario that is supported by the present study. If we consider first an instantaneous configuration of disorder there are extended states, in region C, that are  separated by a mobility edge from localized states, in regions B and A. However due to the dynamic disorder even the localized states can get a finite mobility. For states closest from the mobility edge (region B), this mobility is induced by the dephasing process described in this paper.  Farther from the mobility edge (region C) one expects that the diffusion will be dominated by thermally activated hopping as is usual in disordered systems. We suggest that the small Urbach energy ($E_U \leq 15 meV$) that is measured in MAPbI3 could be related to this region C. The right panel of figure (\ref{Scenario}) shows a qualitative behavior of the mobility $\mu$ and of its static $\mu_S$ and dynamic $\mu_{LD}$ components when disorder increases.

To conclude this work shows that electrons and holes in MAPbI3 have similar transport properties. We showed that the polaronic state is not stable at room temperature and that the dominant effect is that of {Fr\"ohlich} scattering of electrons and holes. We expect that other effects like dipolar field of MA cation or other source of distortion of the lattice are less important. The emerging picture is that of strong quantum localization effects due to the {Fr\"ohlich} scattering or to extrinsic disorder. The quantum localization that we predict in MAPbI3 can appear experimentally similar to a polaronic effect but it is induced by the strong potential disorder. The maximum  mobility at room temperature is  about 200 $cm^2/(V.s)$ and we find that for mobilities below $\mu_{c}\simeq 50 cm^2/(V.s)$ the electronic diffusion is mainly due to the dynamic disorder. In this process states that are localized by the static disorder  can diffuse thanks to the dephasing process induced by the dynamic disorder. This differs from the hopping conduction which is a thermally activated process.  We  suggest that this mechanism of electronic diffusion induced by the lattice dynamics, which has rarely been observed in inorganic semi-conductors, could be observed more frequently in soft materials like for example crystalline organic semi-conductors or hybride perovskites.

We acknowledge fruitful discussions with many colleagues and wish to thank  Jacky Even, Claudine Katan, Paulina Plochocka, Julien Delahaye, Dang Le-Si, and Gabriele Davino. We also thank Ghassen Jemai and Kevin-Davis Richler for their help during this study.

\newpage
\bibliography{biblio}

\newpage
\appendix

\section{Drude-Anderson Model}

\subsection{General relations}
One of the focus of our work has been the computation of the mean squared displacement of charge carrier as a function of time and energy, $X^2(E,t)$. In our formalism this quantity is defined as

\begin{equation}
X^2(E,t)= \frac{Tr([\hat{X}(t)-\hat{X}(0)]^2 \delta(E-\hat{H}))}{Tr(\delta(E-\hat{H}))}.
\end{equation}
We then compute the thermodynamic average over the density of states as
\begin{equation}
X^2(t)= \frac{ \int_0^\infty n(E)e^{-\beta (E - \mu)}X^2(E,t) dE}{\int_0^\infty n(E)e^{-\beta (E - \mu)}dE},
\end{equation}
which gives us the average spread through the lattice of a thermally distributed population of charge carriers, and simply relates to the classical diffusion coefficient $D$ and mobility $\mu$,
\begin{equation}
D = \lim_{t\to\infty} \frac{X^2(t)}{2t},
\end{equation}
\begin{equation}
\mu=\frac{eD}{k_bT}.
\end{equation}
Another value of interest, which behaviour we are able to model more easily is the velocity self correlation function as a function of time, $C(t)$, which is defined as
\begin{equation}
C(t)= \frac{1}{2} \braket{\hat{V_x}(t)\hat{V_x}(0) + \hat{V_x}(0)\hat{V_x}(t)}.
\end{equation}
It represents the correlation that exist between the velocity of a charge carrier at time $t$ and it's initial velocity at $t=0$. This quantity is directly related to the diffusion processes at play in the material. $C(t)$ and $X^2(t)$ are related as 
\begin{equation}
\frac{d X^2(t)}{2dt}=  \int_0^{t} C(t')dt'.
\label{C_t_to_X}
\end{equation}
From them we can the derive formulas for the conductivity, depending on what is more convenient one can use any of the following forms,
\begin{equation}
\sigma(\omega)=  ne^2\frac{tanh(\beta\hbar\omega/2)}{\hbar\omega/2} Re\int_0^{\infty} e^{i\omega t} C(t) dt = ne\mu(\omega),
\end{equation}
\begin{equation}
\sigma(\omega)=  -ne^2\omega^2\frac{tanh(\beta\hbar\omega/2)}{\hbar\omega} Re\int_0^{\infty} e^{i\omega t} X^2(t) dt ,
\end{equation}
The correlation function at time $t=0$ is classically related to the effective mass and the temperature as 
\begin{equation}
C(0) = \braket{V(0)^2}= \frac{k_B T}{m^*} \simeq 2,3.10^{14} cm^2/s^2,
\label{CD}
\end{equation}
for an effective mass $m^*=0.2me$.

\begin{figure}[h]
\includegraphics[width=0.99\linewidth]{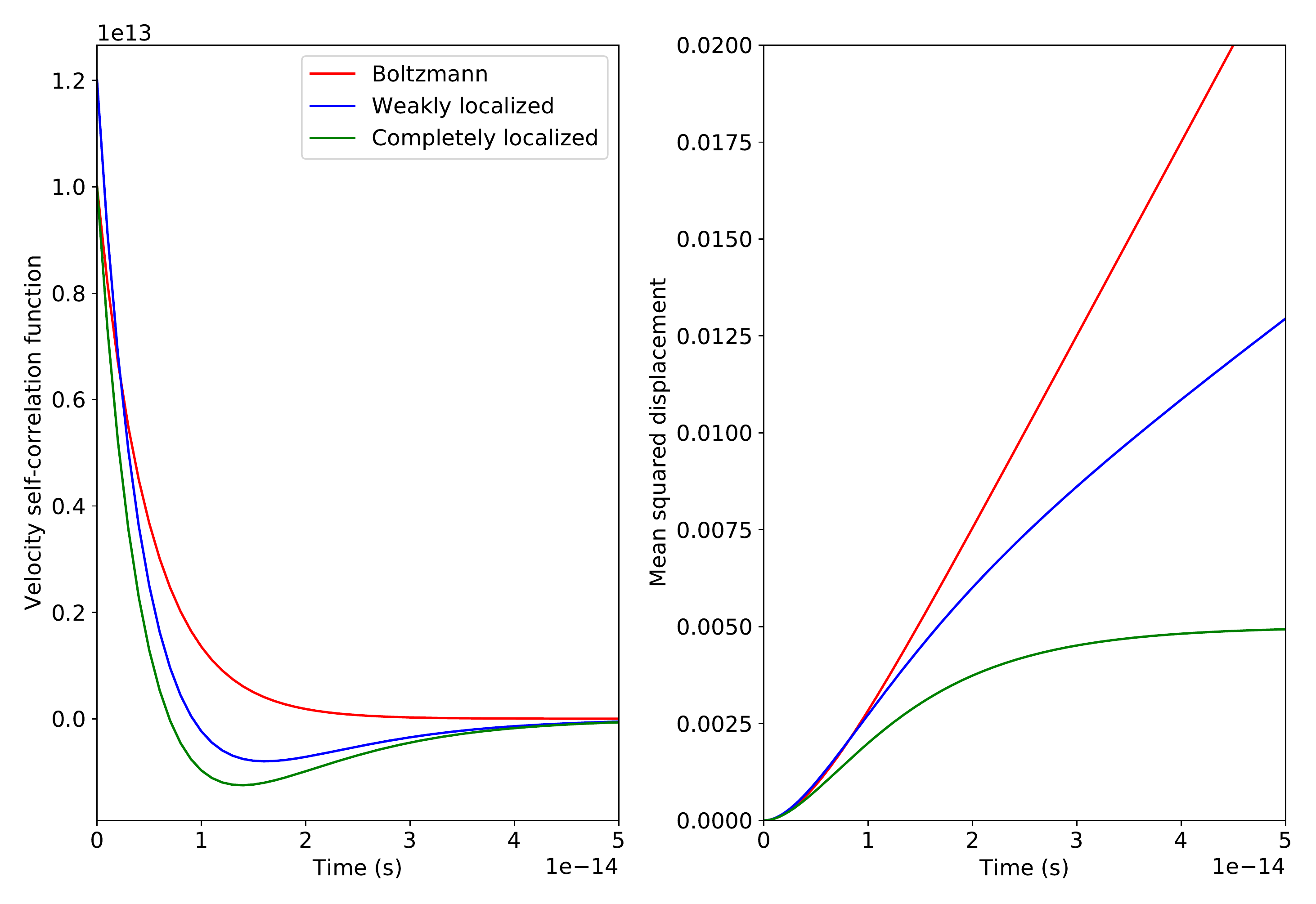}
\caption{(Color online) \textbf{Left:} Velocity self-correlation function as a function of time for different regimes. \textbf{Right:} Corresponding mean squared displacement as function of time.}
\label{fig:autocorellation}
\end{figure}

\subsection{Approximation of the Drude-Anderson model}
The shape of $C(t)$ is directly related to the interaction charge carrier have with the lattice. Drude's model of electronic transport considers non interacting electrons, moving through the material via successive collisions with the much more massive and thus immobile atoms. The average time between collisions is $\tau$ and thus the probability for an electron to undergo a collision is

\begin{equation}
dP= \frac{dt}{\tau}.
\end{equation}

One of Drude's approximation is that, after a collision, electrons completely loose the memory of their initial velocity, thus the velocity after the collision is not correlated to the one before. The evolution over time of self-correlation function of the velocity, $C(t)$, is thus directly proportional to the probability of collision, and we have

\begin{equation}
\frac{dC(t)}{dt} =  C(t)\frac{dP}{dt} = \frac{C(t)}{\tau}.
\end{equation}

and thus 
\begin{equation}
C(t)= Ce^{-t/\tau},
\end{equation}

With $\hat{V_x}(t)= d\hat{X}(t)/dt$. The Drude-Anderson model builds upon the Drude model by adding the possibility for electron to backscatter, creating negative correlation with the original velocity after multiple collisions. This corresponds to backscattering which is at the heart of the Anderson localization phenomenon. Thus appears a negative correlation term for characteristic time $\tau_{B}>\tau_C$:

\begin{equation}
C(t)=  C_Ce^{-t/\tau_C} - C_Be^{-t/\tau_{B}}e^{-t/\tau_{\phi}},
\label{C(t)}
\end{equation}
Where $C_D$ and $\tau_{C}$ relate to the classical diffusion behaviour, and $C_B$ and $\tau_B$ relate to the backscattering phenomenon, and $\tau_{\phi}$ describes the effect the de-phasing processes, in our case the LO phonons, have on localization. These dephasing effects can be seen simply as a renormalization of $\tau_{B}$. 
\subsection{Diffusion and conduction in the Drude-Anderson model}
Combining this form with \ref{C_t_to_X} gives analytical forms to $X^2(t)$, $\mu$, $\sigma(\omega)$, and the localization length $L_{LOC}$:
\begin{equation}
\frac{\Delta X^2(t)}{2} = C_C\tau_C^2(e^{-t/\tau_{C}} - 1 )-  C_B\tau_{B}^2(e^{-t/\tau_{B}}-1) + t(C_C\tau_{C} -C_B\tau_{B}),
\end{equation}

\begin{equation}
L_{LOC}^2 =2 (C_B\tau_B^2- C_C\tau_C^2) = 2 C_B\tau_B^2(1 - \frac{\tau_C}{\tau_B})
\end{equation}

\begin{equation}
\mu = \frac{e}{k_BT} (C_C\tau_{C} -C_B\tau_B)
\end{equation}

\begin{equation}
\sigma(\omega) = ne^2 \frac{tanh(\beta\hbar\omega/2)}{\hbar\omega/2} (\frac{C_C\tau_{C}}{\omega^2\tau_{C}^2 +1 } - \frac{C_B\tau_B}{\omega^2\tau_B^2 +1 }).
\end{equation}

Note that we always have $C_C\tau_{C} \geq C_B\tau_B$ and $\tau_{C} \leq \tau_B$ and as such $\sigma(\omega)$ is always positive. In the case of non localized electrons, $C_B=0$, and in the case of electrons completely localized at long times, we have $C_C\tau_{C} =C_B\tau_{B}$.

\begin{table}
\begin{tabular}{|c|c|c|c|c|}
  \hline
  $\omega_E (meV)$ & $C_c (10^{15}m^2/s^2)$ & $\tau_c (fs)$ & $C_B (10^{15}m^2/s^2)$ & $\tau_B (fs)$ \\
  \hline
  12.5 & 1.23 & 10.3 & 0.25  & 31.4 \\
  10 & 1.82 & 8.28 & 0.7 & 18.3 \\
  7.5 & 3.37 & 2.56 & 1.02  & 8.16 \\
  \hline
\end{tabular}
\caption{Drude-Anderson parameters obtained for the different values of $\omega_e$ studied.}
\label{table_drude}
\end{table}

Surprisingly the Drude model which  totally neglects quantum localization  and the present calculations predict comparable scattering times of the order of $10^{-14}s$ for mobilities in the range of $100 cm^2/Vs$. This is related to the fact that the strong electrostatic disorder has two opposite effects. On one hand the disorder scatters strongly the states and tends to localize them but on the other hand it leads to a strong mixing of states at band edges with states away from band edges which have a much larger average square velocity. This increased average square velocity strongly increases the quantum diffusion at short times. This is shown by the coefficient $C_C \simeq 1.82 . 10^{15} cm^2/s^2$ (see Table \ref{table_drude}), which is much higher than found in the Drude model $C_D \simeq 2,3 . 10^{14} cm^2/s^2$ (see above equation \ref{CD} ), and tends to increase when disorder increases.

\section{Modelization of disorder and Impact of the Methyl-ammonium molecule}
\subsection{Parameter W for the statistical distribution of potential}

We consider first the distribution for electrostatic potential. On a given site $0$ the potential is the sum of potentials created by the various dipoles. 
We note $i=(k,l)$ the global index that represents the site $k$ and the component $l$ of the dipole.

\begin{equation}
V(0)= \sum_{i} \frac{1}{4\pi \epsilon _0 \epsilon_r}  \frac{q_{i}\vec{d_{i}}.\vec{r_{i}}}{r_{i}^3} 
\end{equation}

We define a rescaled  atomic displacement $\vec{D_{i}}=\vec{d_{i}}/\epsilon_r$ then :

\begin{equation}
V(0)= \sum_{i} \frac{1}{4\pi \epsilon _0 }  \frac{q_{i} \vec{D_{i}}.\vec{r_{i}}}{r_{i}^3} 
\end{equation}

Using the stiffness at site $i$ : $k_i=m_i \omega_{e}^2$ the probability distribution for $\vec{D_{i}}$ is 

\begin{equation}
P_{i} \propto  exp(-1/2 k_{i} d_{i}^2/k_{b}T) \propto exp(-1/2 m_{i} \omega_{E}^2 \epsilon_{r}^2 D_{i}^2/k_{b}T) 
\end{equation}

with $W=k_{b}T/(\omega_{E}^2 \epsilon_{r}^2 )$

\begin{equation}
P_{i} \propto exp(-1/2 m_{i} D_{i}^2/W)
\end{equation}

Therefore the probability distribution of the dipoles depends only on the parameter $W=k_{b}T/(\omega_{E}^2 \epsilon_{r}^2 )$.

\subsection{Band gap and parameter $\omega_E$}

The variation of the hopping integrals due to atomic motion is found to be significant, of the order of 20\% on average. Although we find that this disorder has little effect on charge carriers mobility, it has a sizable effect on the band-gap of MAPI and tends to increase it from 1.6eV (i.e. the value for the perfect periodic lattice) to 1.8eV. This increase of the gap is reminiscent of what happens with some distortions or tilts of the octahedrons\cite{Gap_first_principle, Gap_tuning}. This effect tends to be compensated at room temperature by the diagonal disorder. The LO phonons are the source of strong electrostatic potential variation in the material, of the order of $0.5 eV$ and their impact on charge carrier mobilities is considerable. Using the atomic potentials of \cite{tyson} as basis for the motion of atoms, and imposing the band-gap to be 1.6 eV at room temperature, we find that the value of $\omega_E$ for this model is close to 10 meV.

\subsection{Impact of the MA cation on mobilities}

We have studied the impact of the MA molecule and compared it to that of the LO phonon modes. We consider a dipole moment of the MA molecule of $2.1D$ and neglect the correlation in orientation between neighbouring molecules\cite{MA_disorder, MA_disorder2}. In the case of the MA molecule alone, without disorder from the LO phonon, and in the worst case scenario of $\epsilon_r=5$, we found that they limit the charge carrier mobilities from around 900 to 1000 $cm^2/(V.s)$, in concordance with other studies \cite{mu_MA}.

\begin{figure}
\includegraphics[width=0.99\linewidth]{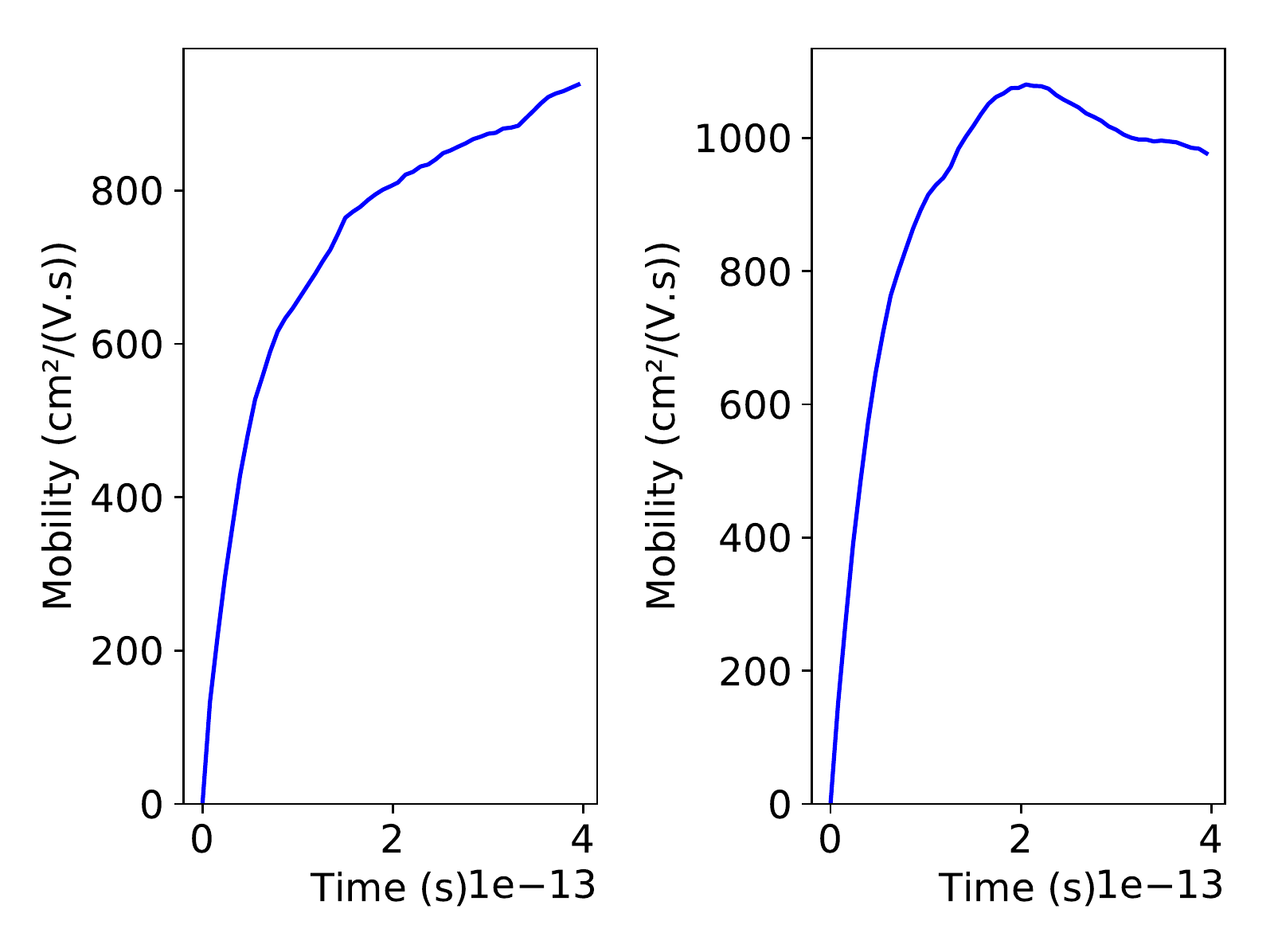}
\caption{ Thermodynamically averaged mobilities as a function of time for holes (left) and electron (right) for a sample containing only MA molecules disorder.}
\label{MA_alone}
\end{figure}
\begin{figure}
\includegraphics[width=0.99\linewidth]{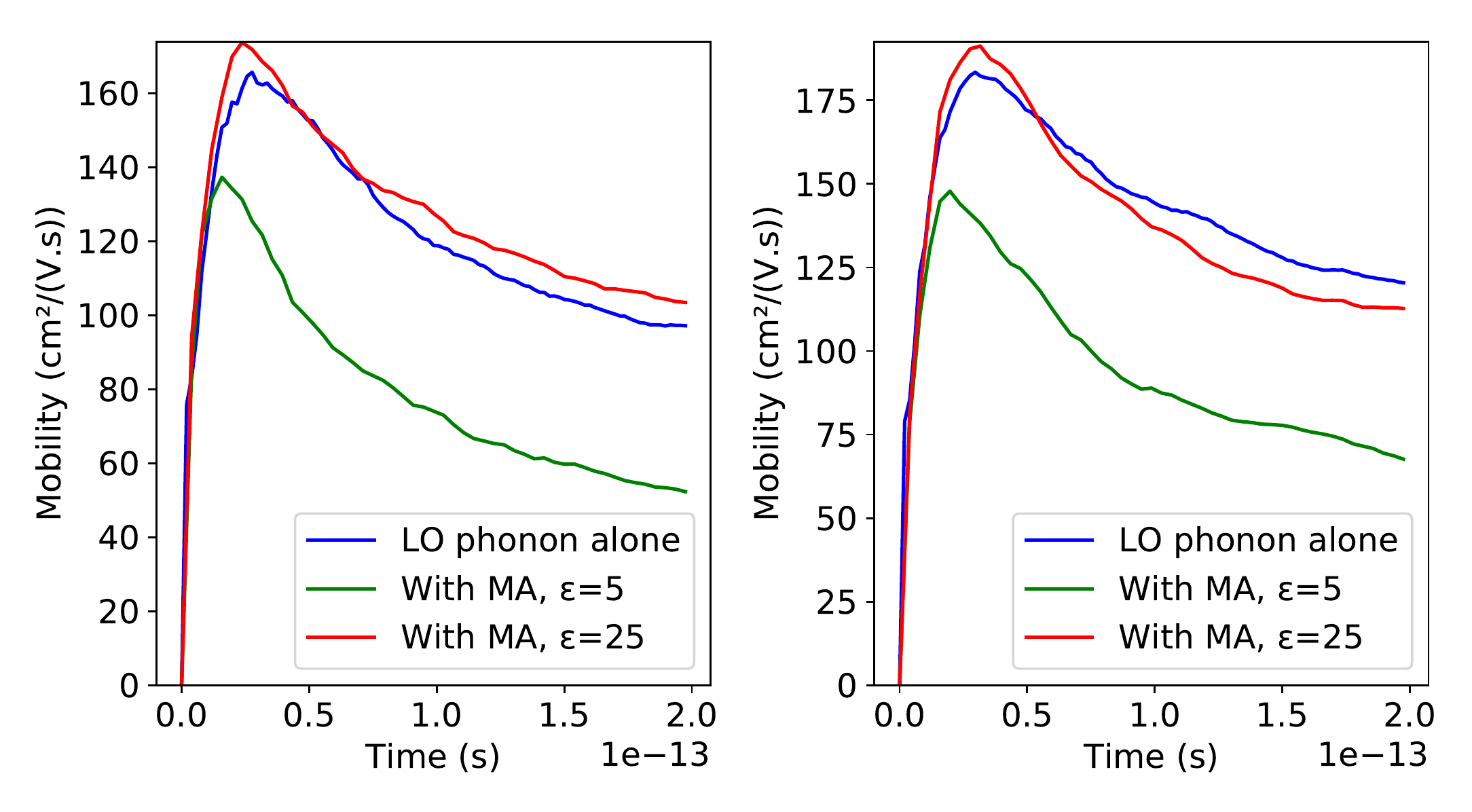}
\caption{(Color online) Thermodynamically averaged mobilities for holes (left) and electrons (right) for different combinations of disorder. Previously shown mobilities results for LO phonons with $\omega_{E}=10$ are shown in blue, and results obtained when adding the MA molecules with different $\epsilon_r$ are shown in green ($\epsilon_r=5$) and red ($\epsilon_r=25$). }
\label{MA_with_LO}
\end{figure}
When looking at the system with both LO phonons and MA molecules, one need to consider the characteristic times of these phenomenon. The LO phonon motion happen on time constants of $0.5ps$ while the MA molecules have much bigger time constant of around $10ps$. As such the MA molecules are screened by the phonons and the dielectric permittivity associated is expected to be much bigger than 5, up to 25. We present both cases of $\epsilon_r=5$ and $\epsilon_r=25$. We find that in the worst case (see Figure \ref{MA_with_LO}), MA molecules have a weak effect on mobilities, and in the most realistic case, have nearly no effect. It is interesting to note however, that in a classical scenario, when adding disorder of the LO phonon with MA, one would expect to have the relations :
\begin{equation}
\frac{1}{\tau} = \frac{1}{\tau_{LO}} + \frac{1}{\tau_{MA}}, 
\end{equation}
\begin{equation}
\frac{1}{\mu} = \frac{1}{\mu_{LO}} + \frac{1}{\mu_{MA}}.
\end{equation}
Considering the mobilities obtained for the disorder separated from one another (see Figures \ref{MA_alone} and \ref{MA_with_LO}), this classical relation predicts mobilities of around $90 cm^2/(V.s)$, which is higher than mobilities of $50 cm^2/(V.s)$ computed in Figure \ref{MA_with_LO}. This reinforces the idea that quantum interferences play a major role in MAPI, and that the addition of even small quantities of disorder to the system tends to quickly localize charge carriers, and the need for a model that goes beyond the semi-classical limit.

Furthermore, the addition of the MA molecules can be seen as a widening of onsite energies distribution. This widening is well fitted by decreasing the LO phonon energy from $\omega_{E}=10mev$ to $9.25meV$. As such, the MA molecules have for effect, at best, a renormalization of $\omega_{E}$ from the point of view of transport. Because we chose the value $\omega_{E}=10meV$ to fit the bandgap of $1.6eV$, we believe our value already encases the effect of the MA molecules.

\end{document}